# Laser Beam Profile Influence on LIBS Analytical Capabilities: Single vs. Multimode Beam

Vasily Lednev[a]*[a], Sergey M. Pershin[a], Alexey F. Bunkin[a]

Single vs. multimode laser beams have been compared for laser ablation on steel samples. Laser plasma properties and analytical capabilities (precision, limit of detection) were used as key parameters for comparison. Peak fluence at focal spot has been observed to be higher for Gaussian beam despite ~14-fold lower pulse energy. A comparison of Gaussian and multimode beams with equal energy was carried out in order to estimate influence of beam profile only. Single mode lasing (Gaussian beam) results in better reproducibility of analytical signals compared to multimode lasing while laser energy reproducibility was the same for both cases. Precision improvements were attributed to more stable laser ablation due to better reproducibility of beam profile fluence at laser spot. Plasma temperature and electron density were higher for Gaussian laser beam. Calibration curves were obtained for four elements under study (Cr, Mn, Si, Cu). Two sampling (drilling and scanning procedures) and two optical detection schemes (side-view and optical fiber) were used to compare Gaussian and multimode beam profile influence on analytical capabilities of LIBS. We have found that multimode beam sampling was strongly influenced by surface effects (impurities, defects etc.). For all sampling and detection schemes, better precision was obtained if Gaussian beam was used for sampling. In case of single-spot sampling better limits of detection were achieved for multimode beam. If laser sources have same wavelength and equal energy than quality of laser beam became a crucial parameter which determined plasma properties and analytical capabilities of LIBS.

## Introduction

Laser induced breakdown spectroscopy (LIBS) is one of the perspective methods for express multi element analysis of samples in different states (solid, liquid, gas)[1,2]. Laser parameters (energy, wavelength, etc.) are strongly influence laser matter interaction and consequently analytical capabilities of method. Influence of laser wavelength[3], laser fluence[4], pulse duration[5] and burst of pulses[6,7] on plasma properties and LIBS analytical capabilities were extensively studied in literature. In most cases a solid state Nd:YAG laser is used for LIBS measurements because such lasers provide a reliable, compact, low price and easy to use source of laser pulses[1,2]. Depending on used Nd:YAG laser model a wide variety of output laser characteristics can be obtained: output wavelength; pulse duration; double pulse mode; beam profile. Beam profile can be different depending on laser model[8]: a Gaussian (single mode, $TEM_{00}$) profile for higher stability and smaller laser spot; multimode ($TEM_{xy}$) profile for higher energy; super-Gaussian profile for higher energy; flat-top profile; "camomile" beam profile[9]. It should be noted that other output laser characteristics are also depend on chosen resonator type (stable, unstable) and lasing regime (single mode or multimode).

Usually, in any application of laser spectroscopy it is preferable to use single mode lasing for

*[a]corresponding author,
e-mail: lednev@kapella.gpi.ru
Wave Research Center at Prokhorov General Physics Institute, Russian Academy of Sciences,
Vavilov. str. 38, Moscow, Russia; Tel: +7 499 538 87 58

Nd:YAG laser since better reproducibility of pulse energy can be achieved. However, such choice of lasing mode is not straightforward for LIBS. If single mode lasing (Gaussian beam profile) is chosen than better reproducibility of laser energy will be achieved and better precision of analytical signal should be obtained. Gaussian beam can be focused into smallest spot compared to other profiles thus higher fluence at focal plane or better spatial resolution in chemical mapping applications can be obtained. On the other hand single mode lasing will result in decrease of the pulse energy thus less mass of sample will be ablated and less energy can be transferred to plasma excitation. It should results in decrease of analytical signal and should reduce the sensitivity of analysis. For laser systems based on single oscillator (low price or compact system) such choice of lasing mode will dramatically determine output laser pulse characeristics (reproducibility; pulse energy and laser spot size: spatial resolution, fluence) and consequently analytical capabilities of LIBS system. To the best of our knowledge, study of optimal lasing regime for LIBS has not been carried out in literature so far.

There are three characteristics of output laser beam that should be substantially different for single mode and multimode lasing: energy, beam profile and reproducibility. Influence of pulse energy (fluence) on laser ablation was systematically studied in literature[1]. A few studies of beam profile influence on laser plasma properties have been also published in literature. All papers were focused on particular features of beam profile influence on laser ablation rather then on influence for analytical capabilities of LIBS. Chalear et al. [10] indicated that stability of analytical signal can be increased if only central part of inhomogeneous multimode laser beam from excimer laser is used. However, authors didn't give any quantitative evaluation of such improvement. Several theoretical and experimental studies were carried out to find out the best beam (Gaussian or "flat-top") for high resolution depth profile analysis by LIBS [11,12]. Comparison of laser ablation with "flat-top" and super-Gaussian beam profiles were carried out by Laserna's group[13]. Plasma properties were compared in terms of plasma temperature and electron density but no impact of beam transformation on analytical figures of merit were discussed. For laser ablation sampling at ICP - MS [14] a multimode beam profile was transformed to the "flat-top" profile in order to improve laser ablation. Better reproducibility of sampling and decrease of fractionation were achieved in this work. It was explained that "flat-top" beam profile resulted in more stable ablation, less droplets was formed and better atomization of sample was obtained. In most recent work concerning beam profile influence a Gaussian and a "spoiled" beams were used for laser ablation in resonant enhanced LIBS[15]. It was observed that Gaussian profile give better reproducibility, signal and longer emission time compared to "spoiled" profile. Better analytical results were also achieved for laser ablation with Gaussian profile than with "spoiled" profile. Finally, reproducibility of laser pulse energy and beam profile should be lower for multimode lasing according laser's theory [8] thus lower reproducibility of laser ablation should be achieved.

The purpose of this study is to compare different lasing regimes (single and multimode) for laser induced breakdown spectroscopy i.e. to evaluate influence of different laser beams (Gaussian and multimode) on laser ablation process and analytical capabilities of LIBS. In this study we have compared influence of beam profile on plasma properties and analytical capabilities of LIBS for three cases of beams: Gaussian beam, multimode beam and multimode beam with energy equal to Gaussian beam. It should be pointed out that in the all previous beam profile studies mentioned above output laser beam were modified after laser output. In presented work beam profile hasn't been modified by any optical system. The different beam profiles were obtained as a result of lasing modes. This study should indicate better design of portable (or low cost) LIBS systems based on single resonator: is it worthy to use single mode lasing and to loose 90% of pulse energy (and ablated mass) in order to increase fluence in laser spot and improve repeatability of laser energy. Steel samples with low alloy additives were used for comparison of LIBS analytical capabilities.

**Experiment**

The presented in Figure 1 experimental setup was used for comparison of different lasing modes. Laser plasma was generated in air by focusing a laser beam normally onto the sample surface. Solid state Nd:YAG laser (1064 nm, 10 ns, 5 Hz) with flash lamp pumping were used to excite the plasma. Laser can be operated in two lasing regimes: single transverse mode lasing (TEM00) and multiple transverse modes lasing ($TEM_{xy}$). Changing of lasing modes was made by diaphragm introduction into laser cavity: single mode lasing with diaphragm (Gaussian beam) and multimode lasing without pinhole (multimode beam). According laser resonator theory[8] maximum diameter should be equal or less 1.6 mm for single mode lasing in our laser system (stable resonator, 350 mm cavity length with plane mirrors and active element rod 6 mm in diameter ). Consequently, a 1.4 mm wide pinhole was used in setup. Laser beam profile measurements were performed with CMOS – camera and neutral optical filters. Beam quality product ($M^2$) were measured according to recommendations of ISO16.CMOS camera were placed in 10 points before and after focal plane and for each position a beam profile was detected. Then beam parameter product ($M^2$) was determined by fitting of second momentum beam width (D4σ) as function of distance from beam waist. The focusing lens, of 90 mm focal length, was placed 89 mm from the sample surface. Exact position of focal plane was determined during $M^2$ measurements.

Two optical detection schemes were used in this setup. First scheme was side-view scheme (Fig. 1b) with quartz lens (F = 120 mm) used for plasma image projection with 1:1 magnification on spectrograph slit. This arrangement allowed to detect space resolved spectra and emission from central part of laser plasma was collected in the present study (dimensions 0.05x4 mm ). Second optical scheme was a scheme with spatial-integrated emission detection. In this case quartz fiber optic was used to collect plasma emission and to transfer it to spectrograph slit. Optical fiber bundle (100 μm diameter) was placed 30 mm from laser spot and under 41$^0$ angle to sample surface that allow to detect the emission coming from all the plasma regions (Fig. 1a). Such optical scheme with quartz waveguide was used only for analytical capabilities comparison. Spectrograph (Andor Shamrock SR – 303i) with gated ICCD (Andor iStar) were used for spectra detection and time resolved measurements. A low noise microphone and oscilloscope were used for optoacoustic measurements. A first minimum in acoustic oscillogram was selected as signal since time delay between laser pulse and first minimum was equal to the time needed for sound wave in air to travel between laser spot and microphone.

Reference samples of low - alloy steel were used for comparison of Gaussian and multimode beam laser sampling. Samples composition is presented in Table 1.

**Table 1.** Elemental composition of reference low-alloy steel samples, wt. %. Elements under analysis are marked bold in the table.

| Sample | C | Si | Mn | P | S | Cr | Ni | Cu | Al | Ti | V | Mo | As | Sn | Pb | Zn |
|---|---|---|---|---|---|---|---|---|---|---|---|---|---|---|---|---|
| sample1 | 0.166 | **0.58** | **1.52** | 0.008 | 0.008 | **0.66** | 0.133 | **0.165** | 0.033 | 0.003 | 0.041 | 0.013 | 0.009 | 0.007 | 0.003 | 0.011 |
| sample2 | 0.328 | **0.67** | **0.96** | 0.018 | 0.020 | **0.038** | 0.060 | **0.059** | 0.005 | 0.0017 | 0.004 | 0.009 | 0.002 | - | - | - |
| sample3 | 0.348 | **1.25** | **0.91** | 0.010 | 0.016 | **1.16** | 0.133 | **0.76** | 0.015 | 0.004 | 0.006 | 0.011 | 0.005 | - | - | - |
| sample4 | 0.105 | **0.30** | **1.63** | 0.007 | 0.004 | **0.101** | 0.093 | **0.184** | 0.039 | 0.023 | 0.082 | 0.010 | 0.007 | 0.009 | 0.005 | 0.010 |
| sample5 | 0.0043 | **0.014** | **0.132** | 0.006 | 0.006 | **0.017** | 0.014 | **0.020** | 0.033 | 0.065 | 0.004 | 0.002 | 0.002 | 0.004 | - | - |

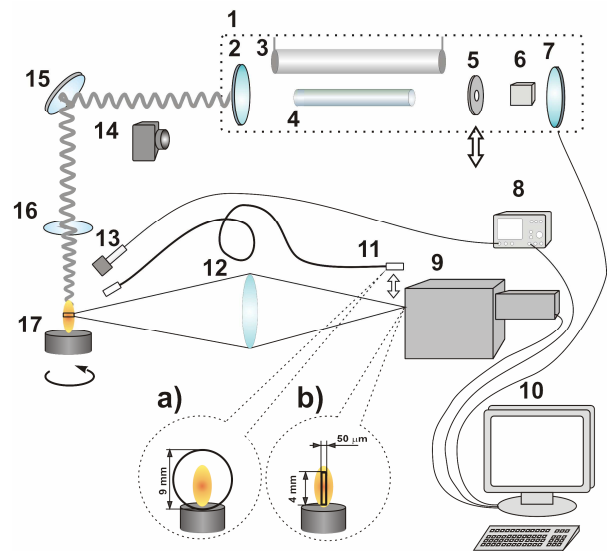

**Figure 1.** Experimental setup

1. Nd:YAG laser (λ = 1064 nm, 6 mJ/pulse < E < 80 mJ/pulse, τ = 10 ns) 2. front mirror, 3. flash - lamp, 4. active element rod (d=5.9 mm), 5. diaphragm (d=1.4 mm), 6. Q-switch, 7. rear mirror, 8. oscilloscope, 9. spectrograph with ICCD, 10. computer, 11. quartz optical fiber, 12. quartz collecting lens (F = 120 mm), 13. microphone, 14. CMOS camera (for beam profile study), 15. mirror, 16. focusing lens (F = 110 mm), 17. sample (a rotation can be used if needed)

Two optical scheme of signal collecting were implemented:

a) scheme with optical fiber detection and spatially integrated signal

b) scheme with side-view detection and spatially resolved signal

In order to increase stability of signal and to diminish influence of impurities at the surface all samples were polished before every measurement with sandpaper (ISO grit designation P 2400 ). Laser crater profiles were measured with white light interferometer microscope (NewView 6200,

Zygo Corp.).

## Results and Discussion

### 1. Laser beam profile in far and near fields

A detailed study of laser beam profile at laser output (near field) and at laser spot on sample surface (far filed) was performed (Table 2). Two lasing regimes (single and multimode) result in two different beam profiles: Gaussian profile for single mode beam ($TEM_{00}$) and multimode profile. Single mode lasing was achieved by diaphragm placement in laser cavity. Output beam diameter in such case was about 1.1 mm. For multimode lasing mode no diaphragm was used and beam with 4.7x4.5 mm dimensions was obtained at laser output (output mirror 2 in Fig.1). Laser pulse energy (measured by radiant power meter Oriel model 70260 with thermopile detector 70261) for single mode beam was observed to be 14 times less compared to multimode beam with the same flash lamp pumping. Measured reproducibility of pulse energy (by energy meter and by photodiode) was nearly the same for both lasing modes (about 2%). Despite 14 times lower energy for laser beam at single mode lasing a peak value of fluence at laser output were higher for Gaussian beam (Fig. 2 a). Beam profile of multimode beam (Fig. 2 c) can be described as complex profile formed by multiple peaks. Beam profile was unstable i.e. peaks position and intensity were fluctuating during pulse – to – pulse study. For 10 successive laser pulses we have detected profiles and than determined fluctuation of beam profile: standard deviation of fluence (coordinate z) was determined for every point at profile (x and y were the same for single z coordinate). Reproducibility of laser beam distribution for Gaussian profile was higher compared to multimode beam profile. Fluctuations for Gaussian beam did not exceed 3% while for multimode beam this parameter was at least 10% (Fig. 2 c and d). However, some peaks at multimode beam were rather stable. These facts are well explained by laser theory of multimode lasing[17,18]: single mode lasing is reproducible while multimode lasing is very unstable. Beam quality for two beams was compared by beam parameter product ($M^2$). Beam quality of single mode beam was slightly poorer compared to ideal Gaussian beam (Table 2) while for multimode lasing mode beam this characteristic was poor. For estimation of beam profile influence only a laser beams with equal energy are needed thus an optical filters were used to decrease energy of original multimode beam. Consequently, beam quality and fluctuations was the same as for multimode beam (with 83 mJ pulse).

**Table 2**. Laser parameters for single and multimode lasing modes (Gaussian and multimode beams)

| Parameter | Single mode laser beam | Multimode 83 mJ (and 6 mJ)[c] laser beam |
|---|---|---|
| **Laser beam (near field):** | | |
| Energy, mJ/pulse | 6 | 83 (6) |
| Energy reproducibility (RSD[a]), % | 1.8 | 1.9 |
| Laser beam profile, | Gaussian, | Multipeak |
| dimensions at 1/10 amplitude, mm | 1.1 x 1.1 | 4.7 x 4.5 |
| Energy density reproducibility[b]: | | |
| average / highest value (RSD), % | 1.4 / 3 | 5 / 14 |
| Beam parameter product, $M^2$ | 5 | 200 |
| **Laser beam spot (far field):** | | |
| Spot dimensions measured by | | |
| CMOS (1/10 amplitude), μm | 110 x 110 | 550 x 500 |
| | | (520 x 490) |
| single shot crater, μm | 120 x 120 | 570 x 510 |
| | | (550 x 500) |
| Energy density[d]: | | |
| at maximum, J/cm2 | 110 | 59 (4.5) |
| CMOS average, J/cm2 | 54 | 31 (2.2) |
| crater average, J/cm2 | 76 | 50 (3.6) |
| Energy density reproducibility | | |
| at target surface: | | |
| average / highest value (RSD), % | 1.8 / 5 | 5.1 / 11 |
| **Crater** | | |
| Crater dimensions after 100 pulses, | 62 x 60 x 8 | 500 x 450 x 6 |
| $l$ x $w$ x $h$, μm | | (490 x 450 x 1.5) |
| Crater volume after 100 pulses, mm³ | $[22 \pm 6]*10^{-6}$ | $[310 \pm 120]*10^{-6}$ |
| | | $([60 \pm 3]*10^{-6})$ |
| Volume of rim after 100 pulses, mm³ | $[30 \pm 10]*10^{-6}$ | $[100 \pm 50]*10^{-6}$ |
| | | $([10 \pm 8]*10^{-6})$ |
| Plasma dimensions, $w$ x $h$ mm | 1.8 x 2.3 | 5 x 8 (2 x 1.6) |
| Optoacoustic signal, mV | 30 ± 1 | 360 ± 6 |
| | | (16 ± 0.5) |

[a] Relative standard deviation

[b] Reproducibility were measured as RSD for each point of beam profile fluence (z) with fixed coordinates (x,y); average is mean value of RSD; highest is a maximum value of RSD (most unstable point at beam profile)

[c] Values for multimode beam with energy equal to Gaussian beam are enclosed in round brackets

[d] Maximum value is a peak value for beam profile fluence, average values were calculated as energy divided by area at laser spot: measured by CMOS camera (at 1/10 amplitude) or by single shot crater area (one shot on steel sample)

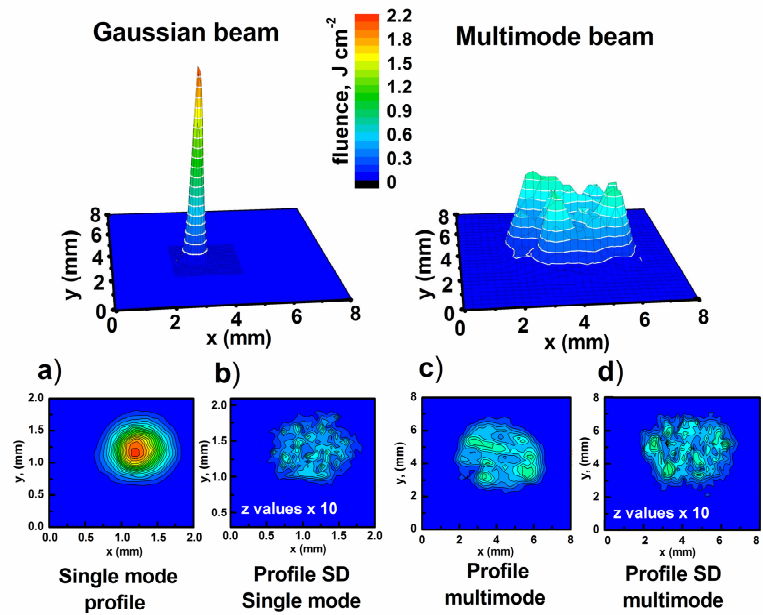

**Figure 2**. Laser beam profiles at near field (at laser output).
Both beam profiles in 3D are presented in equal scale. Figures a) and b) are beam profile and standard deviation of single mode beam profile (Gaussian) (SD were multiplied by factor x10 for better presentation). Figures c) and d) are beam profile and standard deviation of multimode beam. Standard deviation for beam profile is calculated as SD for z-coordinate (fluence) with fixed x,y coordinates for ten successive images of beam profile

In order to determine actual fluence profile at sample surface we have studied a beam profiles at focal spot on sample surface (Fig. 3 ). Single mode beam profile was nearly – Gaussian and exhibits high reproducibility of laser beam profile (RSD <2% ). Multimode beam profile can be described as multimode profile with "flatter" peaks. Peak position and intensity fluctuations were smaller compared to multimode beam at near field (Fig. 2). Reproducibility (RSD) of multimode beam profile was three times poorer compared to Gaussian profile (Fig. 3). Fluctuations of fluence with up to 11% RSD at different local points were observed. For Gaussian (G beam) and multimode (M beam) (83 mJ) beams fluence values were far above ablation threshold[3]. It should be noted that peak value of fluence at sample surface was 2 times higher for Gaussian beam despite 14-fold lower energy at laser output. For the multimode beam with energy equal to Gaussian beam (MeG beam) fluence at the laser spot was 3 times above ablation threshold.

Laser crater profiles measurements were made after 100 successive laser pulses on fresh sample surface for all laser beam profiles (Fig. 4). For first two cases of beams (G and M beams) craters have nearly the same depth but diameters were 10 - fold different. Crater profiles should be described differently: narrow deep crater with smooth walls for Gaussian profile and wide flat crater with ripples on crater bottom for multimode profile. Estimated crater volume formed by Gaussian beam was 14 times smaller compared to crater volume formed by multimode beam (Table 2). For MeG beam crater diameter was nearly the same as for multimode beam with 83 mJ energy but crater depth was 5 times smaller.

It should be mentioned that different forms of rim were determined for craters obtained with different beams. Small rims were observed for craters formed by both multimode beams (6 or 83 mJ). The rim volume formed by single mode laser beam was comparable to crater volume. This fact was attributed to melt splash during laser ablation with Gaussian beam. Craters inner surface profile and shape were different for Gaussian and multimode beams: for single mode beam inner walls and bottom of crater were smooth and craters profiles were nearly the same for replicate measurements; for multimode beam (83 mJ) profile a crater bottom had a pattern with many ripples on surface (~ 20 μm size). Such pattern of crater for multimode beam was reproducible for replicate measurements while local features were different for every crater. For multimode beam with low energy (6 mJ) ripples were also formed at crater bottom. Profile pattern form of multimode crater was attributed to instabilities of multimode beam profile at focal spot. Additionally thick and wide oxide layer was observed for multimode laser beam while for single mode beam almost no oxides were detected after 100 pulses.

Dimensions of laser plasma for different beam sources were detected with CMOS camera (time integrated image). Plasma dimensions and ablated mass comparison for two cases of laser beam (Table 2) lead to supposition that in case of multimode beam sampling density of laser plasma should be lower compared to density of plasma formed with Gaussian beam sampling.

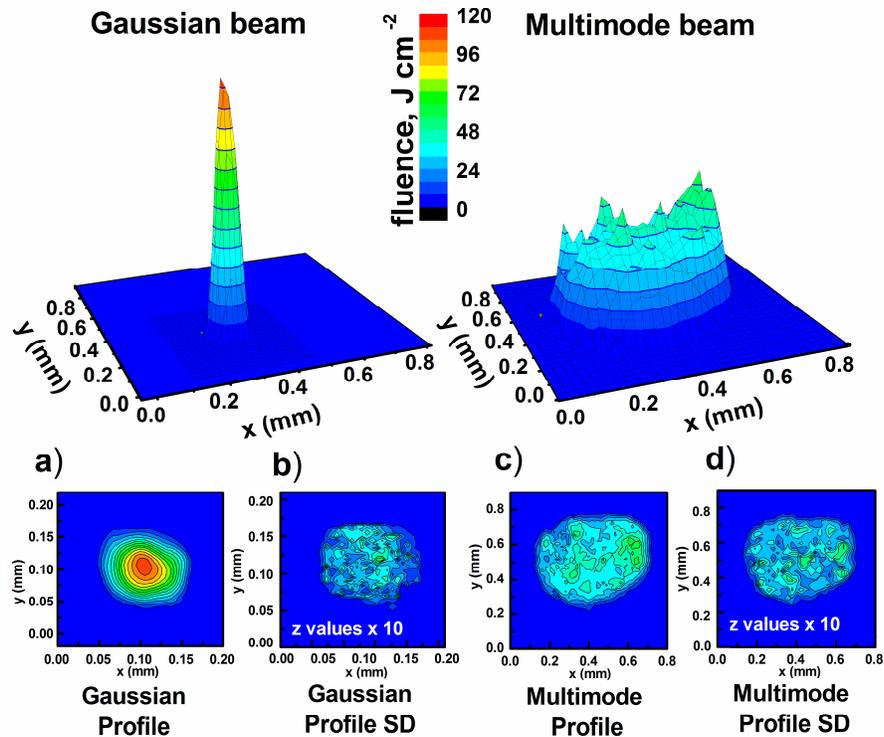

**Figure 3.** Laser beam profiles at target surface (focal plane is 0.8 mm under surface)
Beam profiles in 3D are presented in equal dimension scale. Figures a) and b) corresponds to beam profile and standard deviation of single mode beam profile (SD were multiplied by factor x10 for better presentation). Figures c) and d) are beam profile and standard deviation for multimode beam

## 2. Spectra and Signal

In LIBS, the choice of spectral region and specific analytical lines depends on several factors: spectral interference, transition probabilities, detector sensitivity and possibility of self-absorption. Additionally, in analytical atomic spectrometry, it is conventionally to use an internal standardization by comparing the analytical line intensity with that of the major (matrix) component of the sample. Procedure of internal standardization in LIBS compensates pulse - to - pulse variations in the amount of ablated matter and in the excitation characteristics of plasma. The better choice is to use matrix line which upper level of transition has a similar energy to interested analytical line since lower influence of possible temperature instabilities. Based on above discussed criteria two spectral regions (centered on 280 and 330 nm) were chosen for Cr, Si, Mn, Cu determination. These spectral regions include both atomic and ionic lines of elements under interest and matrix component (Fe). Additionally,
spectroscopic characteristics of chosen lines (atomic lines with low and high energy of upper level; ionic lines) are quite different that allow performing comprehensive comparison of LIBS analytical capabilities for different sources of laser irradiation. A list of analytical lines used for each element and spectroscopic parameters of lines are presented in Table 3.

Laser plasma was obtained with two laser beams and resulted spectra are compared in Figure 5. It was observed that under the same timescale condition (width 1 μs, delay 5 μs) intensity of plasma spectrum for laser plasma created with multimode (83 mJ) laser beam was ~ $10^3$ larger compared to Gaussian beam. However better signal – to – background ratio ($I_{analyt}/I_{background}$) and higher ion – to – atom intensity ratios were achieved for laser plasma spectrum obtained with Gaussian beam at chosen gating condition. For MeG beam plasma lifetime was less 5 μs thus spectrum was detected with another gating (width 1 μs, delay 2 μs).

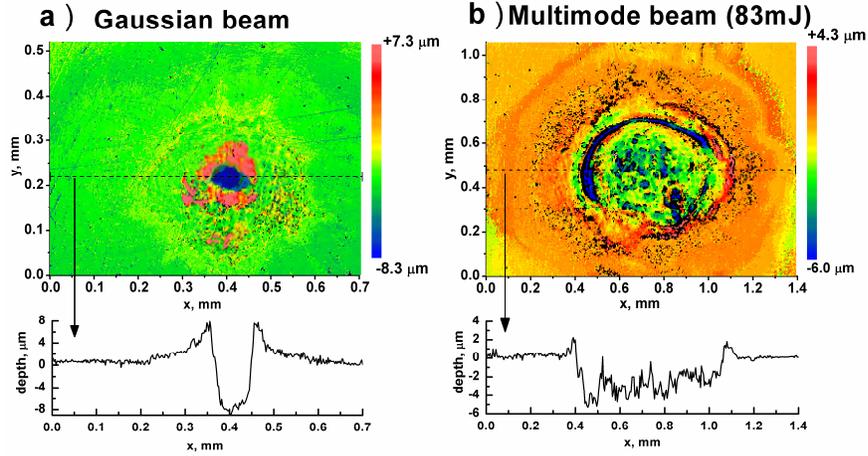

**Figure 4**. Crater profiles after 100 laser pulses for Gaussian laser beam (a) and multimode laser beam (b) profiles. Profiles measured with white light interferometer are presented on top; corresponding 2D profiles are presented on bottom. For multimode beam with low energy (6 mJ) crater (not presented on figure) has nearly the same diameter with crater depth about 1 μm

In order to compare time-integrated emission for two plasmas and evaluate the spectrum that can be obtained with non-gated detectors a series of spectra with different gating parameters were summed and result is presented in Fig.5 b for different cases of laser beams. Time integrated intensity of most strong lines was observed to be ~200 –fold greater and better signal – to – background ratio was achieved for multimode laser beam source (83 mJ). Slightly better spectral resolution for time integrated spectra was achieved for multimode beam (83 mJ) spectra (compare lines of Fe II 276.75 and Fe II 276.92 ) due to different dynamics of laser plasmas. Based on these advantages it can be recommended to use multimode lasing mode for LIBS systems with non gated detector (CCD or Photodiode array for low price or compact systems) since both stronger signal and better signal – to – noise ratio can be achieved. Improved spectrum for multimode beam (83 mJ) should be attributed to greater energy only because for MeG beam we have detected low intensity spectrum with wide unresolved lines.

**Table 3**. Atomic and ionic lines constants from NIST and Kurucz's databases: wavelength, transition probability, degeneracy of upper level, energy of upper level ($E_k$) and energy of lower level ($E_i$). Analytical and matrix lines used for calibration are marked bold

| Wavelength, nm | $A_{ki}*10^7$, s$^{-1}$ | $g_k$ | $E_i$, eV | $E_k$, eV |
|---|---|---|---|---|
| **Fe II 273.07** | **2.5** | **4** | **1.076** | **5.615** |
| **Cr II 283.56** | **20** | **12** | **1.549** | **5.920** |
| **Si I 288.16** | **18.9** | **3** | **0.781** | **5.082** |
| **Mn I 279.48** | **37.0** | **8** | **0.0** | **4.434** |
| **Cu I 324.75** | **13.7** | **4** | **0.0** | **3.816** |
| **Fe I 330.63** | **6.1** | **5** | **2.221** | **5.971** |
| Fe I 370.93 | 1.56 | 7 | 0.915 | 4.256 |
| Fe I 372.76 | 2.25 | 5 | 0.958 | 4.283 |
| Fe I 373.49 | 9.02 | 11 | 0.859 | 4.177 |
| Fe I 374.56 | 1.15 | 7 | 0.087 | 3.396 |
| Fe I 376.55 | 9.8 | 15 | 3.236 | 6.528 |

## 3. Plasma temperature and electron density

Temperature and electron density are important characteristics of plasma in analytical atomic spectroscopy because these are key parameters for atomization and excitation in plasma source. In order to correctly compare sampling with different laser beams we have determined temperature and electron density of plasma. Temperature of laser plasma was determined by Boltzmann plot method with Fe I lines (360 – 375 nm). We used spectral lines with non resonant transition thus the assumption of optically thin plasma was made for the selected lines. Atomic and ionic line constants presented in Table 3 were taken from NIST and Kurtucz's databases[19,20]. Electron density was determined by Stark broadening of Fe I 538.3 line since this line has high value of Stark coefficient which was determined with low experimental error[21]. Line profile was fitted with Voigt function and FWHM value was corrected on instrumental profile. Instrumental profile was estimated by the Fe I 532.8 line with small Stark coefficient at late delay times when electron and ion densities are low and Stark broadening can be assumed negligible.

Results of temperature and electron densities study for various periods of plasma evolution are presented in Figure 6. For Gaussian beam sampling greater temperature and electron density of plasma were observed for first moments. After 5 µsec temperature and electron density was observed to be equal for two cases of laser beam sampling. Higher temperature for first moments of plasma formation can be explained by higher peak fluence for Gaussian beam in laser spot. Fast decay of temperature for Gaussian beam sampling should be attributed to lower ablated mass thus plasma cooling was more fast. For multimode beam with low pulse energy (6 mJ) plasma cools very fast and temperature can be determined only for first 3 µs. Low values of temperature and electron densities for MeG beam case are explained by low fluence at sample surface.

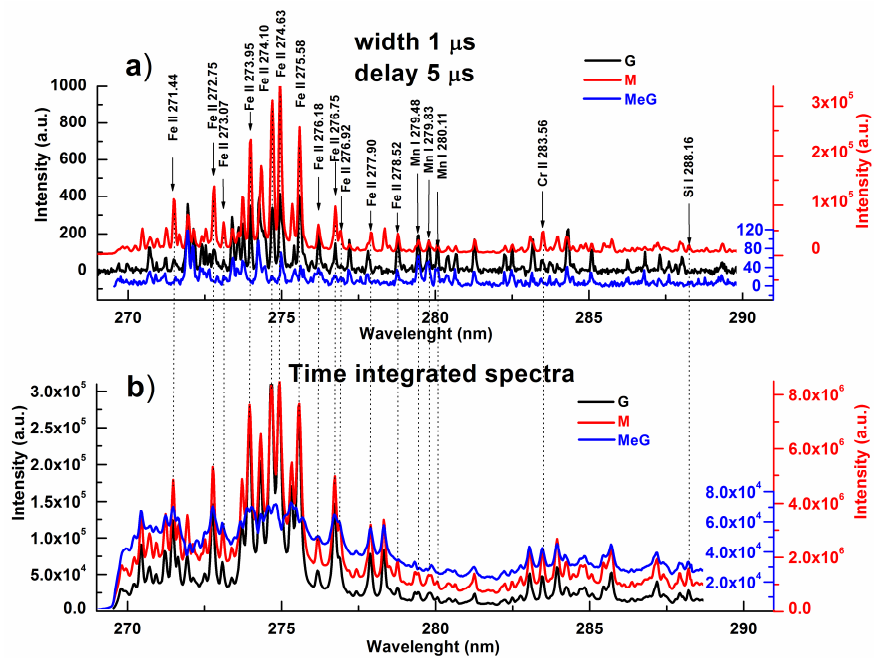

**Figure 5**. Spectra of laser plume obtained with Gaussian (G, black), multimode (M, red) laser beams and multimode beam with energy equal to Gaussian beam (MeG, blue):
a) gated spectra (width 1 µs, delay 5 µs for G and M beams; width 1 µs and delay 2 µs for MeG beam)
b) time integrated spectra (sum of gated spectra during plasma emission)

## 4. Analytical capabilities

The experimental setup was calibrated for four elements (Cu, Si, Mn, Cr) of low-alloy steel samples. Analytical capabilities of different laser sources were compared in terms of precision, limit of detection and regression coefficient of calibration curve. Optimization of signal detection for different beam sampling was performed: sampling procedures, detection schemes and time gated conditions.

*Sampling strategy*

Usually, two procedures of sampling are widely used in LIBS [22,23]: drilling sampling (single-spot sampling) and scanning sampling (multi-spot sampling). First method of sampling uses single-spot strategy with a stationary sample and some pre-pulse treatment before detecting analytical signal. This procedure is used to clear the surface from oxides or contaminations and to increase reproducibility of ablation thus increase precision of signal. However crater formation could influence such sampling with pulse number ascending[24] and preferential evaporation could be significant25. Second procedure implies scanning spot strategy (multiple spots) when every laser shot achieve a new sample surface[26]. Such sampling is achieved by target movement (sample rotation or shifting) so the plasma signal is not influenced by crater formation and by possible non-uniform distribution of sample composition. On the other hand, such sampling is more sensitive to surface effects (contaminations, oxides, mechanical defects) and could suffer from possible instability of lens – to – sample distance during sample movement. Both methods of sampling were used in this study for comparison of laser ablation with different laser beam profiles.

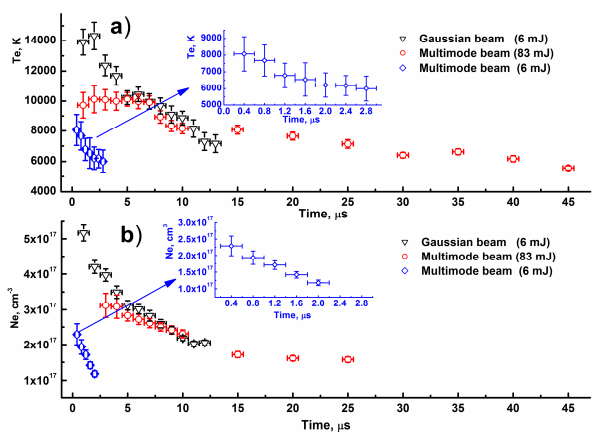

**Figure 6**. Evolution of temperature and electron density for laser plasma formed with different laser beam sources. Temperature and electron density are marked black for Gaussian beam, red and blue for multimode beam (83 and 6 mJ/pulse correspondently)

According to the drilling sampling strategy (single-spot), 500 laser pulses were used for ablation at the same spot to estimate effects of crater formation. We used side-view detection scheme for comparison of sampling strategies. Intensity of matrix (Fe II 273.08) and analytical line (Cr II 283.56) was measured in combination with optoacoustic signal and results are presented in Figure 7. For ablation with Gaussian beam, it was observed that during first 30 laser pulses the intensity of ionic lines reached maximum and then slowly decreased. Mean intensity values didn't change significantly after 100 pulses and pulse – to – pulse intensity reproducibility was constant. For multimode beam ablation (83 mJ), same tendency for increasing of mean intensity during first 50 pulses was detected. However pulse – to – pulse fluctuations were increasing dramatically with pulse number and no stabilization of pulse – to – pulse reproducibility was observed. For multimode beam with low energy (6 mJ), mean value of intensity didn't change significantly for first 100 pulses and signal reproducibility was poorer than for multimode beam with 83 mJ.

We have observed that optoacoustic signal changed during the first 10 pulses for all laser beams. Optoacoustic signal should be proportional to ablated mass as was discussed in paper[10]. This supposition was verified in our experiment conditions for both laser sources (see supplementary materials fig. s1). Consequently, observed reproducibility of ablated mass were rather high and was nearly the same any case of laser beams. After 20 pulses, reproducibility of measured sound signal was more than 6 times better than measured reproducibility of Fe II or Cr II intensity and only 2 times lower than the stability of laser pulse energy. For MeG beam reproducibility was slightly lower compared to other beams.

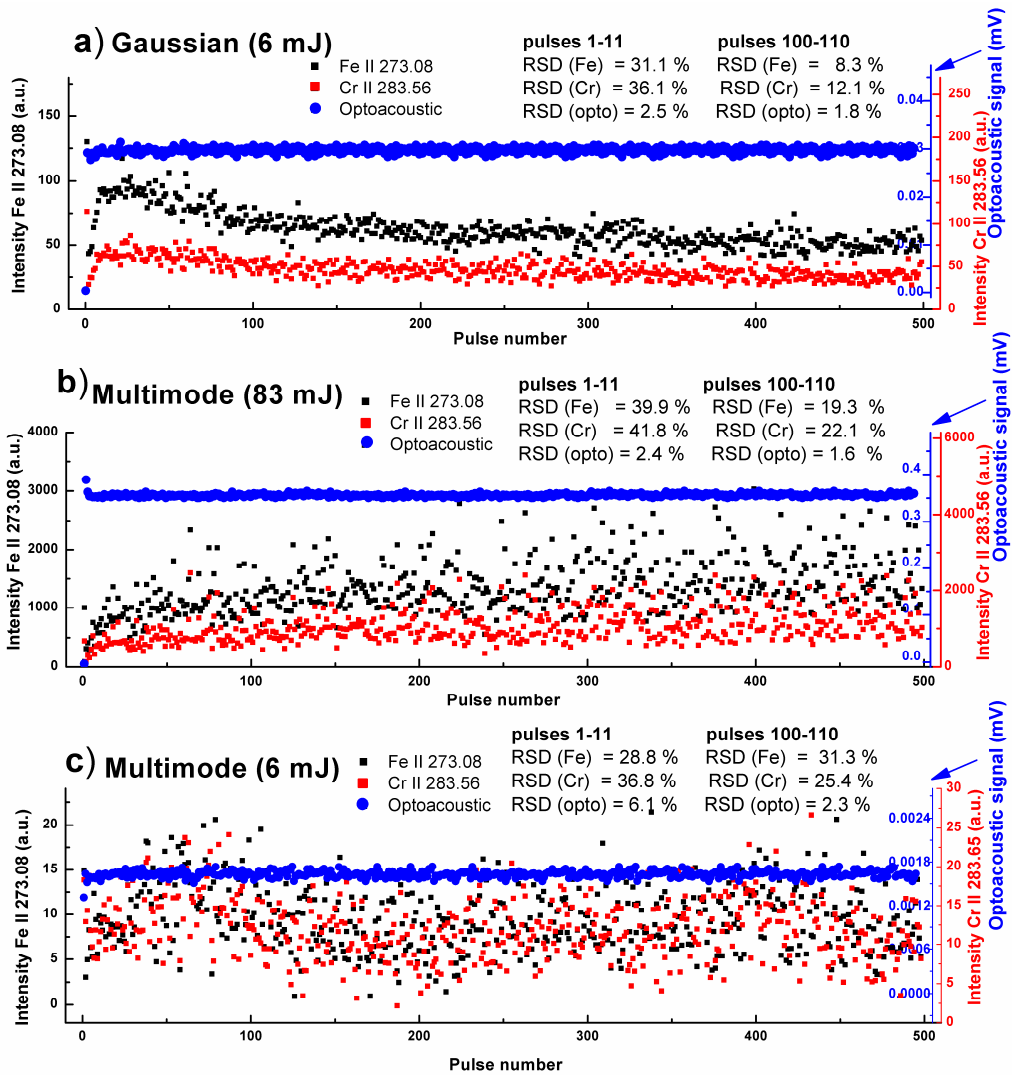

**Figure 7**. Pulse - to - pulse study for single-spot sampling with Gaussian beam (a), for multimode beams for 83 mJ (b) and for 6 mJ (c) sampling.
Intensity of Fe II 273.08 (a.u.) and Cr II 283.56 lines (a.u.) and optoacoustic signal (mV) were detected simultaneously. Gating parameters for spectra detection was width 2 µsec and delay 5 µsec for Gaussian and Multimode (83 mJ); width 2 µsec and delay 1 µsec for Multimode (6 mJ). For optoacoustic signal a first minimum of oscilloscope curve were used as a signal since its time delay were equal to the time that needed for sound to travel from laser spot to microphone.

**Table 4**. Comparison of signal reproducibility for different sampling procedures and plasma characteristics comparison for Gaussian (G beam), multimode 83 mJ (M beam) and multimode 6 mJ (MeG beam) beams.

| Parameter | G beam | M beam | GeM beam |
|---|---|---|---|
| **Sampling procedures**[a] | | | |
| Reproducibility of line intensity for Fe II 273.08 and Cr II 283.56, (RSD), % | | | |
| 1. single-spot sampling: | | | |
|   a) pulse – to – pulse after 100 prepulses | 8 (Fe), 12 (Cr) | 19 (Fe), 22 (Cr) | 31 (Fe), 25 (Cr) |
|   b) averaged by summing of 50 pulses (after 100 prepulses) | 3 (Fe), 6 (Cr) | 6 (Fe), 7 (Cr) | 8 (Fe), 9 (Cr) |
| 2. scanning sampling: | | | |
|   a) pulse - to - pulse | 25 (Fe), 27 (Cr) | 30 (Fe), 31 (Cr) | 36 (Fe), 41 (Cr) |
|   b) average by 50 pulses | 9 (Fe), 11 (Cr) | 14 (Fe), 22 (Cr) | 16 (Fe), 25 (Cr) |
| Optoacoustic signal / error (RSD), mV / % | | | |
|   a) pulse – to – pulse single-spot sampling after 100 pre-pulses | 0.029 / 1.8 | 0.35 / 2.0 | 0.016 / 2.3 |
|   b) multispot sampling | 0.005 / 8.1 | 0.09 / 9.2 | 0.012 / 12.2 |
| Cleaning pulses, number | 30 | 70 | 30 |
| **Plasma properties**: | | | |
| Plasma lifetime [b] | 14 | 50 | 5 |
| Temperature and electron density | | | |
|   Te, K | 14000 - 8000 | 10000 - 5000 | 8000 - 6000 |
|   Ne, 1/cm3 | $(5.1 – 1.9) \cdot 10^{17}$ | $(3.8 - 1.5) \cdot 10^{17}$ | $(2.3 - 1.2) \cdot 10^{17}$ |
| Gating parameters used for calibration curve, µs | gate 20 delay 2 | gate 10 delay 5 | gate 5 delay 0.5 |

[a] Side-view detection scheme was used for comparison

[b] Period of time when strong matrix lines can be detected with signal – to – noise ratio greater than 3

For Gaussian beam profile ablated mass didn't change significantly after 30 pulses and small decrease of intensity should be attributed to crater formation that resulted in lower fluence at crater surface. For case of multimode laser beam ablation we have detected that the mean value of intensity was increasing with pulse number accending while pulse – to – pulse intensity precision was decreasing. Reproducibility of ablated mass was the same for two beams and only one parameter differs dramatically for different laser beams: fluence profile reproducibility is almost 3 times poorer for multimode laser beam compared to Gaussian. Thus decrease of pulse – to – pulse reproducibility of intensity should be attributed to greater instability of fluence at focal plane for multimode laser beam. Additionally, unstable fluence profile resulted in ripples formation of crater bottom (Fig. 4) and next laser pulse of multimode beam with unpredictable fluence profile will interact with such crater that will enhance instability of fluence at crater surface and result in self unstable ablation. This consequence of laser beam interaction with sample (unstable fluence and ripples at crater bottom) will lead to decrease of pulse – to – pulse precision with pulse number ascending that was observed in Figure 7.

Scanning sampling (multiple spots strategy) was achieved by rotation of the sample with every laser pulse arrives at new surface. Reproducibility of signals for all laser beams were poorer compared to stationary target (Table 4). For multimode (6 and 83 mJ) and Gaussian laser beam sources reproducibility of intensities were nearly the same (about 30 %). If multimode beam was used for ablation than the absence of correlation between element concentration and intensity were detected for some analyte lines (Fig. 8 e, f). For analytical lines with high excitation energy correlation could be obtained. The higher excitation energy level the better correlation between intensity and concentration was obtained. Gaussian beam ablation give better results, calibration curve can be obtained for all elements but precision and sensitivity was not so good as was obtained (Fig. 8 a) for stationary sample. Low sensitivity and precision for calibration curve and even absence of correlation between signal and concentration for multimode beam sampling can be explained by strong influence of sample surface (oxides, impurities, etc.) and instabilities of ablation process. Fluctuation of lens – to – sample distance should be neglected in our conditions since this parameter instability was estimated to be less than 0.05 mm. According crater study discussed above higher ratio of diameter – to – depth was obtained for multimode beam thus less material was ablated from bulk compared to Gaussian beam ablation. For multimode beam profile surface influence was dominant under used experimental conditions. Higher peak fluence detected for Gaussian beam profile lead to narrow crater formation and more material is ablated from bulk than from surface.

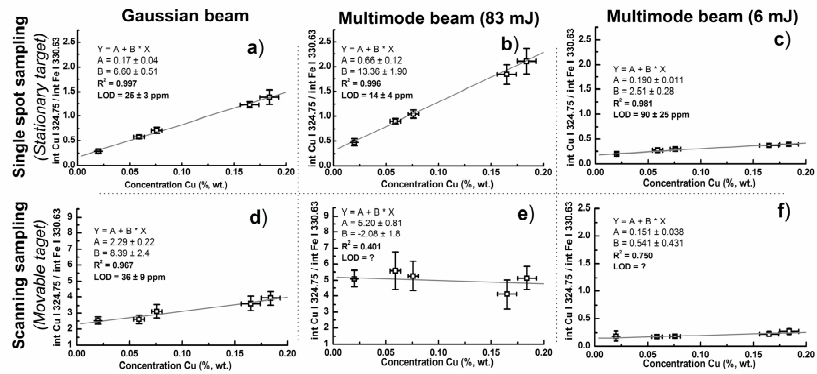

**Figure 8**. Calibration curve of copper for laser sampling with Gaussian and multimode beams.
a) – single-spot sampling with Gaussian beam; b) – single-spot sampling with multimode beam (83 mJ/pulse); c) – single-spot sampling with multimode beam (6 mJ/pulse); d) – scanning sampling with Gaussian beam; e) – scanning sampling with multimode beam (83 mJ/pulse); f) – scanning sampling with multimode beam (6 mJ/pulse)

Based on these results it can be recommended to use single mode lasing for LIBS system based on single resonator (compact or low-cost system) or for single shot analysis (stand-off analysis or analysis of movable objects).

In order to increase reproducibility of obtained data each spectrum was collected by summing of 50 laser pulses. For single-spot sampling a 100 pre-ablation pulses were used before every measurement (to clear sample surface and to obtain stable laser ablation). This procedure resulted in 2-fold better reproducibility of measured intensity RSD (6%) for single and 3-fold better RSD (7%) for multimode beam (Table 4). Same procedure of spectra detection were used for scanning sampling that increase signal reproducibility for up to 3 times for Gaussian and multimode ( 6 and 83 mJ) beams.

Single-spot sampling (stationary target) was used for all experiments with laser plasma study (comparison of signals, reproducibility, plasma temperature etc.) and determination of optimal parameter for analysis

Two detection schemes are widely use in LIBS for spectra registration: spatially resolved scheme (different plasma regions can be studied) and spatially integrated scheme (irradiation from different plasma parts are averaged). Widely used first scheme or side-view scheme (Fig.1 a), implies that plasma image is projected by optical system on spectrograph input (spectrograph entrance slit). This allows detection of emission from different local points of plasma by moving of collecting optics. We used this side-view scheme for all measurement presented above. Second scheme uses optical system that transfers plasma irradiation on detection system with no spatial resolution (backscattering scheme, systems with optical fiber or stand-off telescopes). In our experiments we used a quartz optical fiber for space integrated detection scheme. Second detection optical scheme was used for the following reason. Laser plasma is a source with huge gradient of material density, plasma temperature and electron density. Consequently fluctuations of intensities observed for side-view scheme can be lowered if space integrated scheme is used for spectra detection. Side-view scheme for spectra detection could also lead to the overestimation of signal instability because spectrum is detected from small slice of plasma image ($h \times w$, $4 \times 0.05$ mm) while ripple features size is only 2 times smaller. Based on supposition that different optical schemes could influence on analytical capabilities of the system we compared two optical schemes in our study.

Two detection schemes and two sampling procedures result in four possible ways of signal detection. Only three of these detection combinations were used in this study: single-spot sampling with side-view and optical fiber detection schemes; scanning sampling with side-view. Procedure with scanning sampling and optical fiber detection was not presented in this paper since for multimode beam sampling we have detected absence of correlation between line intensity and concentration for some elements. If Gaussian beam were used for scanning sampling and optical fiber detection scheme than calibration can be performed but sensitivity was too poor for any reasonable analytical measurements.

*Gating parameters*

With laser-induced plasmas, it is generally required to use time delay prior spectra detection in order to avoid the intense initial continuum emission and improve the line resolution. This allows to detect spectra with good spectral resolution, low background and sufficiently high intensity. Optimal gating parameters for calibration were determined for single-spot sampling and spatially resolved optical scheme (side view scheme that used lens projection of laser plasma image with 1:1 magnification). Gating parameters were determined for the three cases of laser beam separately because plasma's dynamics and properties were different. Exposure (width) and delay times were chosen based on better relation signal/(noise+background) (Fig. 9) and linear dynamic range of detector (data presented for Cr II 283.56). For atomic lines (Si, Mn, Cu) nearly the same dependence were observed with small shift of signal – to – background ratio maximum to later detection time. Determined optimal gating parameters (Table 4) were different for different laser beam sources while were chosen same for calibration with different spectral region, selected sampling procedure or chosen detection scheme.

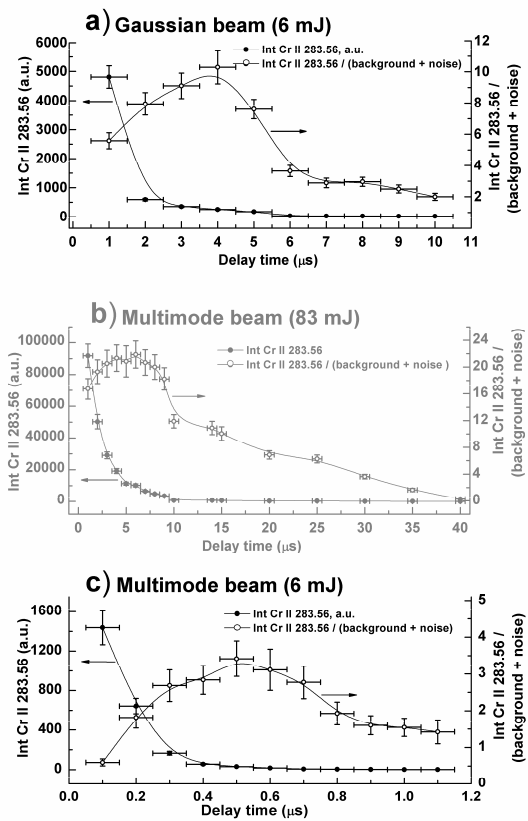

**Figure 9**. Gating optimization for different laser sources: Gaussian beam (a), Multimode beam 83 mJ (b) and Multimode beam 6 mJ (c).

*Calibration curves*

Experiment setup was calibrated on four elements under study: Si, Cr, Mn, Cu. Internal normalization on matrix component were used to eliminate any unwanted experimental fluctuation: intensities of Si I 288.16, Cr II 283.56 and Mn I 279.48 were normalized on Fe II 273.08 line; Cu I 324.75 was normalized on Fe I 330.63 line.

The sampling procedure was described above for single-spot and scanning sampling with summing of 50 pulses. For both sampling strategies a six replicate measurements of spectra were detected for statistics. For single-spot sampling it should diminish possible inhomogeneous distribution of analyte in the sample. For every spectral line, the intensity was obtained by subtracting a background from the intensity of the line. Then intensity of analytic line was normalized on matrix line. The obtained calibration curve were fitted with linear function. The vertical error bars on calibration curve show the standard deviation ratio calculated from the six replicate measurements. The horizontal error bars show concentration error stated for reference samples. Limits of detection (LOD) were calculated with 3σ criteria as recommended by IUPAC[27]: *LOD=3σ/s*, where σ is a standard deviation for the background for sample with lowest analyte content, *s* – sensitivity. Precision for every calibration curve was estimated as mean relative standard deviation for all points at plot: *RSD = [Σ$_i$RSD(y$_i$)]/N*, where *RSD(y$_i$)* – relative standard deviation of normalized intensity for *i*-point on calibration curve, N - number of points.

Table 5 summarizes the results of analytical capabilities comparison for two laser beam sources. Calibration curves for chromium are presented in Figure 10 for different sampling procedures and detection schemes. Detection limits for ionic lines were poor with values exceeded 200 ppm while for atomic resonant lines detection limits were about 20 ppm.

Single-spot sampling with side-view scheme detection resulted in high precision of signal and better limits of detection. Increase of reproducibility was detected for all analyte lines if Gaussian beam was used for sampling. Improved precision was explained by more stable laser ablation with Gaussian beam profile that has been discussed in details above in the text. Multimode beam (83 mJ) sampling improved regression coefficients and gave better LODs for all elements. Detection limit improvement for multimode beam ablation compared to Gaussian beam ablation was observed to depend on spectral line characteristics. The improvement of LOD for multimode beam sampling was increasing if analytical line excitation energy was decreasing. Such dependence should be explained by lower temperature and increase of the ablated mass. For Chromium line lower plasma temperature result in lower ion - to - atom ratio thus smaller concentration of ions were obtained in plasma and hence smallest improvement of LOD were achieved. If multimode beam with pulse energy equal to Gaussian beam was used for sampling (6 mJ) than analytical figures of merit degrades substantially. This fact was explained by low fluence in laser spot and consequently lower ablation mass, lower temperature and incomplete atomization of ablated material in the plasma. Consequently, better limits of detection for multimode beam with 83 mJ energy compared to Gaussian beam should be attributed only to greater energy of laser beam.

It should be pointed out that LODs achieved for the Gaussian beam sampling were slightly poorer compared to the multimode beam (83 mJ) sampling while total laser energy for single mode beam was 14 times lower. This fact is encouraging for development of portable LIBS devices especially in case of micro chip lasers [28,29] which are very promising laser sources for compact LIBS systems.

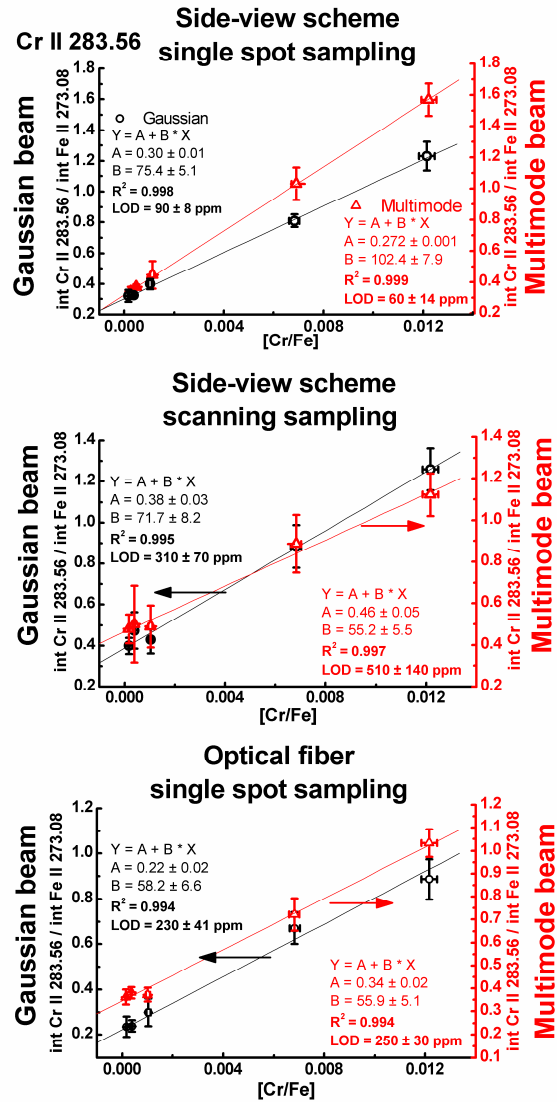

**Figure 10**. Calibration curves of chromium for Gaussian and multimode laser beam sources at different sampling strategies and detection schemes: a) single-spot sampling with side-view scheme; b) scanning sampling with side-view detection scheme; c) single-spot sampling with spatially integrated scheme.

For scanning sampling and side-view scheme detection the substantially higher fluctuation of signal was obtained for all laser beams. Reproducibility of signals decreased two times that in combination with low sensitivity summarized in poor detection limits for all elements. For multimode sampling (6 and 83 mJ) calibration curve was obtained only for chromium line. For other elements no correlation between normalized intensity and concentration were determined. Absence of correlation for multimode sampling was explained by surface influence of laser ablation that was discussed above in details. For Gaussian beam sampling decrease of precision should be attributed to surface influence and instability of lens - to – sample distance.

**Table 5**. Comparison of analytical figures of merit for Gaussian and multimode beam laser ablation: Gaussian beam (G), multimode beam (M) and multimode beam with energy equal to Gaussian beam (MeG); concentration range, wt. %; correlation coefficient, $R^2$; limit of detection (LOD), ppm; precision of signal (relative standard deviation, RSD, %). For cases with absence of correlation between normalized intensity and concentration dash "-" symbol is used in the table. Precision presented in the table was estimated as mean relative standard deviation for all points at plot: $RSD = [\Sigma_i RSD(y_i)]/N$, where $RSD(y_i)$ – error (relative standard deviation) of normalized intensity of $i$-point on calibration curve, $N$ - number of point on calibration plot. Analytical lines are sorted in row by upper level energy of transition for better view (from highest for Cr II 283.56 to lowest for Cu I 324.75)

| experiment (detection scheme, sampling procedure) | | Element | concentration range, wt. % | Laser beam | $R^2$ | LOD, ppm | RSD, % |
|---|---|---|---|---|---|---|---|
| side-view scheme | single-spot sampling | Cr II 283.56 | 0.017 – 0.66 | G | 0.998 | 90 ± 8 | 6.1 |
| | | | | M | 0.999 | 60 ± 14 | 9.2 |
| | | | | MeG | 0.995 | 170 ± 50 | 12.1 |
| | | Si I 288.16 | 0.014 – 1.25 | G | 0.966 | 80 ± 9 | 7.1 |
| | | | | M | 0.999 | 50 ± 8 | 11.3 |
| | | | | MeG | 0.941 | 120 ± 20 | 14.1 |
| | | Mn I 279.48 | 0.132 -1.63 | G | 0.954 | 200 ± 14 | 4.9 |
| | | | | M | 0.969 | 110 ± 20 | 10.1 |
| | | | | MeG | 0.966 | 900 ± 200 | 12.9 |
| | | Cu I 324.75 | 0.020 – 0.76 | G | 0.998 | 25 ± 3 | 5.1 |
| | | | | M | 0.996 | 14 ± 4 | 10.8 |
| | | | | MeG | 0.981 | 90 ± 25 | 12.8 |
| | scanning sampling | Cr II 283.56 | 0.017 – 0.66 | G | 0.995 | 310 ± 70 | 12.8 |
| | | | | M | 0.997 | 510 ± 140 | 19.1 |
| | | | | MeG | 0.885 | 700 ± 200 | 18.4 |
| | | Si I 288.16 | 0.014 – 1.25 | G | 0.934 | 110 ± 30 | 16.1 |
| | | | | M | - | - | - |
| | | | | MeG | - | - | - |
| | | Mn I 279.48 | 0.132 -1.63 | G | 0.973 | 600 ± 100 | 11.8 |
| | | | | M | - | - | - |
| | | | | MeG | - | - | - |
| | | Cu I 324.75 | 0.020 – 0.76 | G | 0.969 | 30 ± 5 | 11.2 |
| | | | | M | - | - | - |
| | | | | MeG | - | - | - |
| spatial integrated scheme | single-spot sampling | Cr II 283.56 | 0.017 – 0.66 | G | 0.994 | 230 ± 41 | 8.9 |
| | | | | M | 0.994 | 250 ± 30 | 8.3 |
| | | Si I 288.16 | 0.014 – 1.25 | G | 0.991 | 190 ± 25 | 9.5 |
| | | | | M | 0.998 | 170 ± 20 | 8.4 |
| | | Mn I 279.48 | 0.132 -1.63 | G | 0.999 | 230 ± 30 | 7.1 |
| | | | | M | 0.991 | 118 ± 20 | 6.1 |
| | | Cu I 324.75 | 0.020 – 0.76 | G | 0.999 | 41 ± 5 | 9.9 |
| | | | | M | 0.991 | 16 ± 3 | 12.1 |

Quartz optical fiber was used for detection with alternative space integrated optical scheme. Intensity in absolute values was $10^3$ lower for such detection scheme since small diameter of fiber. If multimode beam with low energy (6 mJ) was used for sampling than spectrum with small intensity was obtained and this fact didn't allow to use such beam for fiber detection scheme. Reproducibility of signals was same for both types of laser beams. Precision for multimode beam (83 mJ) sampling was improved compared to side-view detection scheme. This fact was attributed to lower fluctuation of space integrated plasma emission compared to local fluctuation in side-view scheme. Signal decrease was the main reason for poorer limits of detection compared to side-view detection. Sensitivity was the same for both cases of laser beams thus analytical capabilities for fiber optics detection scheme were comparable for two different laser sources.

**Conclusions**

Comparison of laser ablation with Gaussian (TEM00) and multimode laser beams generated at single resonator system were carried out. It was observed that despite 14 times lower energy for single mode beam a higher peak fluence for Gaussian beam can be achieved at focal spot. If Gaussian and multimode beams have equal energy than 20 times different fluence profiles was obtained. For multimode beam we have detected that fluence profile was very unstable at focal spot compared to Gaussian beam. Gated and time integrated spectra were compared for two types of laser beams and based on this comparison a multimode beam sampling should be recommended for single resonator (compact or low-cost) LIBS systems.

Higher temperature and electron density were detected for laser plasma created with Gaussian beam that was explained by higher peak fluence at laser spot.

Two sampling procedures (drilling and scanning) were used for comparison of signal precision for Gaussian and multimode beam sampling. It was determined for single – spot sampling that reproducibility of analytical signal is strongly affected by chosen laser beam and this fact was explained by instabilities of fluence profile at focal spot for multimode beam. Scanning sampling resulted in poorer reproducibility for both beams. For multimode beam source it was determined that signal was strongly influenced by surface effects (impurities, etc.) and for some element calibration curve couldn't be obtained for scanning sampling. Consequently, sampling method (single-spot or scanning) should be carefully chosen in case of multimode beam sampling. According these results for LIBS system based on single resonator (compact or low-cost system) it is preferable to use single mode lasing if only one shot at sample surface is possible to achieve (stand-off analysis or analysis of movable objects). Single mode laser beam is also preferable for analysis because of better precision can be achieved. However multimode laser beam should be recommended to use for analysis of trace elements because of higher intensity of spectrum. Better lateral resolution was observed for Gaussian beam and in combination with high reproduction of crater formation this laser source should be recommended to use for depth profile study or chemical mapping applications.

Analytical performance comparison was carried out for four elements under the study (Cr, Cu, Si, Mn) with two sampling procedures and two detection schemes. For any sampling or detection scheme we have observed that better precision was achieved if Gaussian beam were used for the sampling. For all calibration curves a better linearity (characterized by $R^2$) was obtained for the multimode beam sampling. Better sensitivity and limits of detection were achieved if multimode beam was used as a laser source in case of single-spot sampling. For the multimode beam with energy equal to Gaussian beam a degrading of analytical capabilities was observed. Thus better sensitivity obtained for multimode beam with energy 83 mJ was attributed only to higher energy of this beam compared to Gaussian beam. In case of two laser beams have same wavelength and equal energy quality of beam profile became a crucial characteristic that determined plasma properties and analytical capabilities of LIBS.

# Acknowledgements

This work was performed under support Russian Foundation of Basic Research 08-02-00008-a and 09-02-01173-a. Authors would like to thank A.V. Gudilin (WRC GPI), G.N. Vishnyakov (VNIIOFI) and Moiseev N.N. (VNIIOFI) for help with white light interferometry study.